\documentstyle[twoside,fleqn,espcrc2,epsf,bm]{article}

\newcommand{\beq}{\begin{equation}}
\newcommand{\eeq}{\end{equation}}
\newcommand{\bea}{\begin{eqnarray}}
\newcommand{\eea}{\end{eqnarray}}

\newcommand{\order}[1]{{\cal O}(#1)}  

\title{ 
Progress Calculating Decay Constants with NRQCD and AsqTad Actions
} 

\author{
Matthew Wingate,\address{Department of Physics, The Ohio State University,
	Columbus, Ohio 43210, USA}\thanks{Address as of Sep 2003:
Institute for Nuclear Theory, University of Washington, Seattle, WA
98195-1550, USA}
Christine Davies,\address{Department of Physics \& Astronomy,
University of Glasgow, Glasgow, G12 8QQ, UK}
Alan Gray,${}^{\rm b}$
Emel Gulez,${}^{\rm a}$
G.\ Peter Lepage,\address{Laboratory of Elementary Particle Physics,
Cornell University, Ithaca, NY 14853, USA}
and Junko Shigemitsu${}^{\rm a}$
} 
\begin{document}
\pagestyle{empty}

\begin{abstract}
We combine a light AsqTad antiquark with a
nonrelativistic heavy quark to compute the decay constants of 
heavy-light pseudoscalar mesons using the ensemble of 3-flavor 
gauge field configurations generated by the MILC collaboration.  
Preliminary results for $f_{B_s}$ and $f_{D_s}$ are given
and status of the chiral extrapolation to $f_B$ is reported.
We also touch upon results of the perturbative calculation which matches 
matrix elements in the effective theory to the full theory at 1-loop order.
\end{abstract}

\maketitle

\section{INTRODUCTION}

The ``AsqTad'' improved staggered quark action has made feasible 
the simulation of QCD with 3 flavors of sea quarks, with 2 of 
the flavors varying in mass from $m_s$ to below $m_s/5$.  
Unquenched simulation with this action at light sea quark masses
produces agreement with experiment for several quantities: the $\Upsilon$
spectrum, $B$ masses, and $\pi$ and $K$ decay constants.  Using
a single set of input parameters, these quantities could not previously
reproduce experiment \cite{Davies:2003ik}.

Having removed this discrepancy and constructed actions with good 
scaling properties, the uncertainties arising from chiral extrapolations 
can be studied more accurately.  

In recent work \cite{Wingate:2002fh} we proposed and tested the
use of improved staggered quarks as the light quark in heavy-light
mesons.  Since the heavy quark does not have the doubling problem,
taste-breaking effects present in light staggered hadrons are 
suppressed in heavy-light mesons.  Consequently the same operators
used in Wilson fermion simulations can be used here, employing the
identity between naive and staggered quark propagators.

This work utilizes a subset of the public MILC configurations, 
the parameters of which we summarize in Table~\ref{tab:simul};
further details appear in \cite{Bernard:2001av}.  The correct
experimental kaon mass is obtained by tuning the valence strange
quark mass to $au_0 m_q=0.040(1)$
on the $au_0m_{ud}^{\rm sea}=0.01$ lattice.  For clarity, we will quote
quark masses in units of the mass, $m_s$, corresponding to the 
physical strange mass.

The NRQCD (improved through $1/M_0^2$) and improved staggered 
actions are exactly as in \cite{Wingate:2002fh} (see references therein).
The operator matching has been carried out through $\order{1/M_0}$
at 1-loop order in perturbation theory \cite{Morningstar:1997ep,Gulez:2003}.

\begin{table*}[t]
\setlength{\tabcolsep}{1.5pc}
\newlength{\digitwidth} \settowidth{\digitwidth}{\rm 0}
\catcode`?=\active \def?{\kern\digitwidth}
\caption{\label{tab:simul}Simulation parameters and inverse
lattice spacing from two $\Upsilon$ splittings \cite{Gray:2003}.
The lattice volume is $20^3\times 64$.}
\begin{tabular*}{\textwidth}{@{}c@{\extracolsep{\fill}}cccc}
$au_0 m_{ud}^{\rm sea}$ & $au_0 m_s^{\rm sea}$ & $N_{\rm conf}$
& $a^{-1}(2S-1S)$ & $a^{-1}(1P-1S)$ \\ \hline
{0.01} & {0.05} & 568 & {1.59(2)} GeV& 1.58(3) GeV \\
{0.02} & {0.05} & 468 & {1.61(2)} GeV& 1.64(2) GeV\\
{0.03} & {0.05} & 564 & {1.60(3)} GeV& 1.68(4) GeV\\
\hline
\end{tabular*}
\end{table*}

\section{$\bm{f_{B_s}}$ AND $\bm{f_{D_s}}$}

The $B_s$ decay constant is the simplest for us to compute since
no extrapolations in valence quark mass are necessary and dynamical
quark mass dependence is found to be small (see discussion below). 
The strange and bottom quark masses are fixed by tuning the 
bare masses to the physical $K$ and $\Upsilon$ masses.  
On the lattice with $m_{ud}^{\rm sea}\approx m_s/4$ we find 
\beq
f_{B_s} ~=~ 260 \pm 7 \pm 26 \pm 8 \pm 5 ~{\rm MeV} \,.
\label{eq:fbsresult}
\eeq 
The first uncertainty is the statistical errors in the matrix elements
and lattice spacing.  The dominant uncertainty is the estimate of
$\order{\alpha_s^2}$ effects neglected in the 1-loop matching
calculation.  The last two uncertainties estimate the errors due
to higher orders in the heavy quark expansion and discretization errors.

\begin{figure}[t]
\vspace{6cm}
\includegraphics{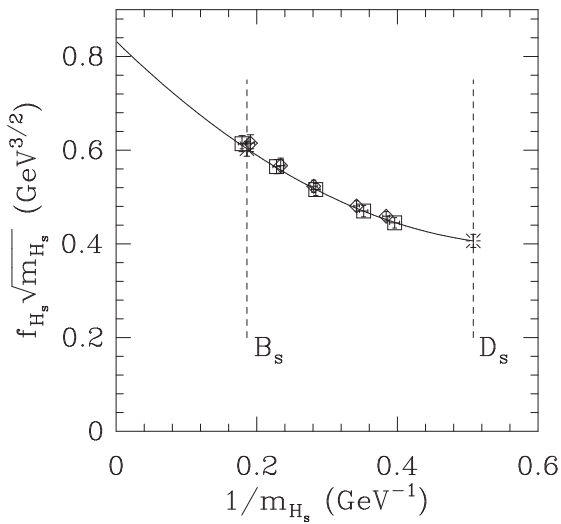}
\caption{\label{fig:PhiHs_vs_invhl} $f_{H_s}\sqrt{m_{H_s}}$ vs.\ inverse
meson mass.  Squares correspond to the light sea quark mass 
$m_{ud}^{\rm sea}\approx m_s/4$ and diamonds to 
$m_{ud}^{\rm sea}\approx m_s/2$.  The curve is the fit (to the squares)
described in the text and the bursts are the fit results at physical
values of $m_{H_s}$.}
\end{figure}

Figure \ref{fig:PhiHs_vs_invhl} shows the mass dependence of the
combination $f_{H_s}\sqrt{m_{H_s}}$.  We would like to extrapolate
to the charm region, where we are unable to calculate using NRQCD
at this lattice spacing.  Fits to a power series
\beq
f_{H_s}\sqrt{m_{H_s}} ~=~ \Phi^{\rm stat}\left( 1 
\;+ \; \sum_{n=1}^N \frac{C_n}{m_{H_s}^n} \right)
\eeq
give acceptable $\chi^2$'s for $N\ge 2$ -- a linear fit gives an
unacceptably large $\chi^2$/DoF = 3.  The extrapolated value
for $f_{D_s}$ is 290 MeV.  The largest uncertainty is again
$\order{\alpha_s^2}\approx 10\%$ perturbative corrections.
Other uncertainties are still being estimated.

\begin{figure}[t]
\vspace{4.3cm}
\includegraphics{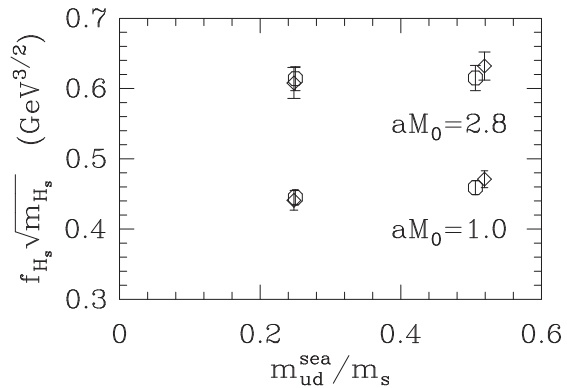}
\caption{\label{fig:PhiHs_vs_rsea} Light sea quark mass dependence
on $f_{H_s}\sqrt{m_{H_s}}$. The heavy quark mass is $\approx m_b$ for
the top 4 points and $\approx 1.5 \,m_c$ for the bottom 4 points.
Octagons use $a^{-1}$ from
$\Upsilon(\mathrm{2S-1S})$ and diamonds use $a^{-1}$ 
from $\Upsilon(\mathrm{1P-1S})$. }
\end{figure}

Within statistical errors, there is no sea quark mass dependence
apparent in Fig.~\ref{fig:PhiHs_vs_invhl}.  In Fig.~\ref{fig:PhiHs_vs_rsea}
we plot the same quantity for 2 values of heavy quark mass and with
2 definitions of the lattice spacing.  Note from Table \ref{tab:simul}
that with $m_{ud}^{\rm sea}\approx (3/4) m_s$
we see the reappearance of a scale ambiguity.  
With $m_{ud}^{\rm sea}\approx m_s/2$ the 2 lattice spacings differ by
$1.5\sigma$.  This effect apparently masks any sea quark mass dependence,
as can be seen in Fig.~\ref{fig:PhiHs_vs_rsea}, consequently we
conclude that a chiral extrapolation in $m_{ud}^{\rm sea}$ will be an
effect smaller than the other quoted errors.  We emphasize that
within statistical errors of 1.3\% no scale ambiguity exists
for $m_{ud}^{\rm sea}\approx m_s/4$ which is where our result
(\ref{eq:fbsresult}) is taken.

\section{CHIRAL BEHAVIOR}

The main benefit of using staggered fermions is being able to
simulate closer to the chiral limit.  In order to make maximal
use out of the gauge field configurations, several values of 
valence quark mass $m_q$ are used.  So far we have accumulated
data with $m_{ud}^{\rm sea} \ge m_s/4$ and $m_q \ge m_s/8$, compared
to $m_q\ge 0.7\,m_s$ which is the state-of-the-art using Wilson-like
fermions \cite{Aoki:2003xb}.

\begin{figure}[t]
\vspace{5cm}
\includegraphics{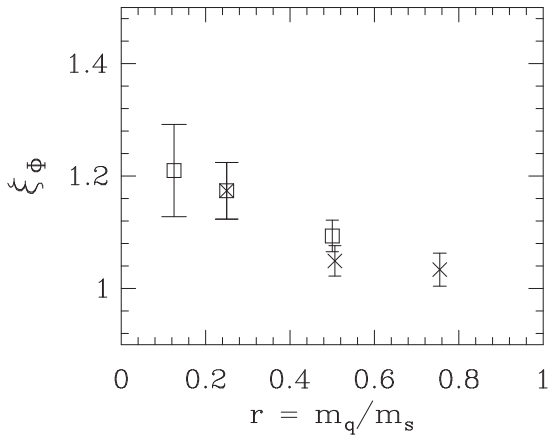}
\caption{\label{fig:xi_PhiB_r} $\xi_\Phi\equiv 
f_{B_s}\sqrt{m_{B_s}}/f_{B_q}\sqrt{m_{B_q}}$ plotted as a function of 
valence light quark mass in units of $m_s$.  The crosses have
$m_q=m_{ud}^{\rm sea}$ and are uncorrelated, and the squares have 
fixed $m_{ud}^{\rm sea}=m_s/4$ so are correlated. }
\end{figure}

In Fig.\ \ref{fig:xi_PhiB_r} we plot $f_{B_s}/f_{B_q}$ (times
$\sqrt{m_{B_s}/m_{B_q}}$ or $1.01$) against $m_q/m_s$.  Crosses are
unquenched, except that the dynamical strange quark mass is
slightly heavier ($m^{\rm sea}=(5/4)m_s$) than the valence 
strange quark mass ($m^{\rm val}=m_s$), and squares
have light sea quark mass fixed to $m_s/4$.  A partially quenched 
analysis will utilize the correlations between data points computed
on the same configurations.

\begin{figure}[t]
\vspace{5cm}
\includegraphics{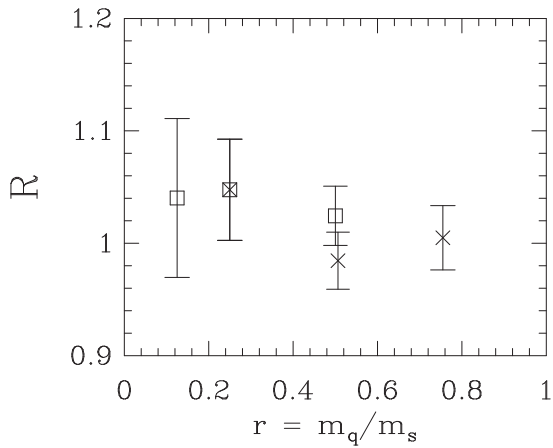}
\caption{\label{fig:R_M28_r} The double ratio $R$ (see Eq.\ (\ref{eq:R}))
plotted as a function of valence light quark mass in units of $m_s$.  
Symbols are as in the previous figure. }
\end{figure}

Be\'cirevi\'c {\it et al.} recently noticed that the coefficient of
the chiral logs in the double ratio
\beq
R ~\equiv~ \frac{f_{B_s}\sqrt{m_{B_s}}}{f_{B}\sqrt{m_{B}}}
\bigg/ \frac{f_K}{f_\pi}
\label{eq:R}
\eeq
is numerically smaller than in either ratio individually 
\cite{Becirevic:2002mh}.  Combining our results with the
MILC collaboration results \cite{Bernard:LAT03}
for $f_K/f_\pi$ (which is 1.22 experimentally)
yields Fig.~\ref{fig:R_M28_r}.  Extrapolating $R$ to $r=0$,
a range of 1.0 -- 1.1 would correspond to $\xi$ between 1.22 and 1.34.
($\sqrt{B_{B_s}/B_{B_d}}= 1.01(3)$ \cite{Aoki:2003xb}.)  Much work
remains to be done before we have a trustworthy and precise final
result, but these simulations with masses below $m_s/2$ should cast
new light on the chiral extrapolation of this important quantity.

\section*{ACKNOWLEDGMENTS}

We are grateful for the availability of the MILC gauge field configurations.
Our simulations were run at NERSC.  This work was supported by the DOE
and PPARC.


\end{document}